\documentclass[prl,superscriptaddress,showpacs,twocolumn]{revtex4}
\usepackage{graphicx}
\begin{document}

\title{Vortex rectification effects in films with periodic asymmetric pinning}

\author{J. Van de Vondel}
\affiliation{Nanoscale Superconductivity and Magnetism Group,
Laboratory for Solid State Physics and Magnetism, K. U. Leuven,
Celestijnenlaan 200 D, B-3001 Leuven, Belgium}

\author{C. C. de Souza Silva}
\affiliation{Nanoscale Superconductivity and Magnetism Group,
Laboratory for Solid State Physics and Magnetism, K. U. Leuven,
Celestijnenlaan 200 D, B-3001 Leuven, Belgium}

\author{B. Y. Zhu}
\affiliation{Quantum Phenomena Observation Technology Laboratory,
The Institute of Physical and Chemical Research(RIKEN) and
Advanced Research Laboratory Hitachi Ltd. Hatoyama, Saitama
350-0395, Japan}

\author{M. Morelle}
\affiliation{Nanoscale Superconductivity and Magnetism Group,
Laboratory for Solid State Physics and Magnetism, K. U. Leuven,
Celestijnenlaan 200 D, B-3001 Leuven, Belgium}

\author{V. V. Moshchalkov}
\email{Victor.Moshchalkov@fys.kuleuven.ac.be}
\affiliation{Nanoscale Superconductivity and Magnetism Group,
Laboratory for Solid State Physics and Magnetism, K. U. Leuven,
Celestijnenlaan 200 D, B-3001 Leuven, Belgium}

\begin{abstract}
We study the transport of vortices excited by an ac current in an
Al film with an array of nanoengineered asymmetric antidots. The
vortex response to the ac current is investigated by detailed
measurements of the voltage output as a function of ac current
amplitude, magnetic field and temperature. The measurements
revealed pronounced voltage rectification effects which are mainly
characterized by the two critical depinning forces of the
asymmetric potential. The shape of the net dc voltage as a
function of the excitation amplitude indicates that our vortex
ratchet behaves in a way very different from standard overdamped
models. Rather, as demonstrated by the observed output signal, the
repinning force, necessary to stop vortex motion, is considerably
smaller than the depinning force, resembling the behavior of the
so-called inertia ratchets. Calculations based on an underdamped
ratchet model provide a very good fit to the experimental data.
\end{abstract}


\maketitle

From the point of view of classical thermodynamics, it is not
possible to induce directed motion of particles by using
equilibrium fluctuations only, otherwise it would constitute a
{\it perpetuum mobile} of the second kind~\cite{SecondLaw}.
Nevertheless, non-equilibrium fluctuations, such as periodic
excitations  or a ``colored" noise, are allowed to take advantage
of the asymmetry of a periodic ratchet potential to promote motion
of particles in a preferential direction~\cite{Reimann_Rev}. New
solid-state-based ratchet systems are currently being developed
for controlling the motion of electrons~\cite{Linke98} and
fluxons, as well as for particle separation~\cite{Matthias03} and
electrophoresis~\cite{Marquet02}. In particular, ratchet
potentials in superconducting devices may be very useful to
control the dissipative motion of fluxons, which causes undesired
internal noise.

Modern lithographic technics make it possible to fabricate
periodic arrays of vortex pinning sites with size and shape that
can be easily tuned, thus giving an interesting perspective for
making different asymmetric pinning potentials. In this context,
several ideas to control flux motion by applying an ac excitation
have been proposed~\cite{Lee99,Wambaugh99,Olson01,Zhu}, but up to
now only a few experiments have been
realized~\cite{Wordenweber04,Villegas03}. One realization has been
recently implemented on a Nb film with a square array of
nanoscopic triangular magnetic dots~\cite{Villegas03}. The authors
reported rectification of the ac driven vortices due to the
asymmetric shape of the dots. Nevertheless, the detailed dynamics
of vortices in such structures is not yet completely understood.

In this Letter we investigate a composite square array of pinning
sites, with its unit cell consisting of a small and a big square
antidot separated by a narrow superconducting wall, as a vortex
rectifier. As demonstrated by our dc and ac transport measurements
at several fields and temperatures, this configuration is able to
break the reflection symmetry of the total effective pinning
potential and promote flux quanta rectification. Moreover, our
data reveals a remarkable hysteresis in the current-induced
pinning-depinning process. This gives an apparent inertia to the
system with important consequences in the overall dynamics.
Finally, we discuss how standard overdamped models fail in
describing these results and propose an underdamped ratchet model
that provides a very good fit to our experimental data.

\begin{figure}[b]
\centering
\includegraphics*[width=8.5cm]{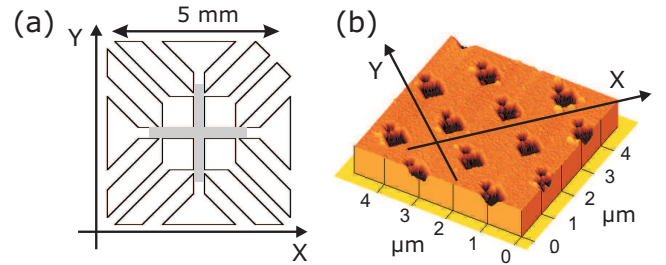}
\caption{Layout of the Al film. (a) Cross-shaped geometry of the
sample to allow for transport measurements in the $x$ and $y$
directions. (b) atomic force micrograph of a $5 \times 5$
$\mu$m$^{2}$ area of the asymmetric pinning sites\label{sample}}
\end{figure}

The pattern was prepared by electron-beam lithography on a SiO$_2$
substrate. The superconducting thin film is a 38 nm thick Al, in a
$5 \times 5$ mm$^{2}$ cross-shaped geometry [see Fig.
\ref{sample}.(a)] to allow four-point electrical transport
measurements in two perpendicular current directions. The central
part of the cross consists of two $300$-$\mu$m-wide strips
containing the nanoengineered array (period $a_p=1.5$ $\mu$m) of
asymmetric pinning sites [see Fig. \ref{sample}.(b)]. This gives a
value of 0.92 mT for the first matching field, $H_1=\Phi_0/a_p^2$
(here $\Phi_0=h/2e$ is the flux quantum). In each strip, the
voltage contacts are 2 mm apart. The sizes of the small and big
antidots are $300 \times\ 300 \ nm^2$ and $600 \times\ 600 \ nm^2$
with a separation between them of 90 nm. The magnetic field was
applied perpendicularly to the sample surface. The superconducting
critical temperature ($T_c$ = 1.469 K, $\triangle T = 3 mK$) was
obtained by using a resistance criterion of 10\% of the normal
state resistance. From the residual resistivity at 4.2 K, the
elastic mean free path ($l_{el}$ = 3.9 nm) was found. From this we
can calculate the coherence length $\xi(0) = 67.5 \ nm $ and the
penetration length $\lambda(0) = 195 \ nm$ using the dirty limit
expressions. We can therefore conclude that the patterned film
with the composite array of square antidots is a type-II
superconductor ($\kappa=\frac{\lambda(0)}{\xi(0}  = 2.89 >
\frac{1}{\sqrt{2}}=0.71$).

\begin{figure}[tb]
\centering
\includegraphics*[width=8.5cm]{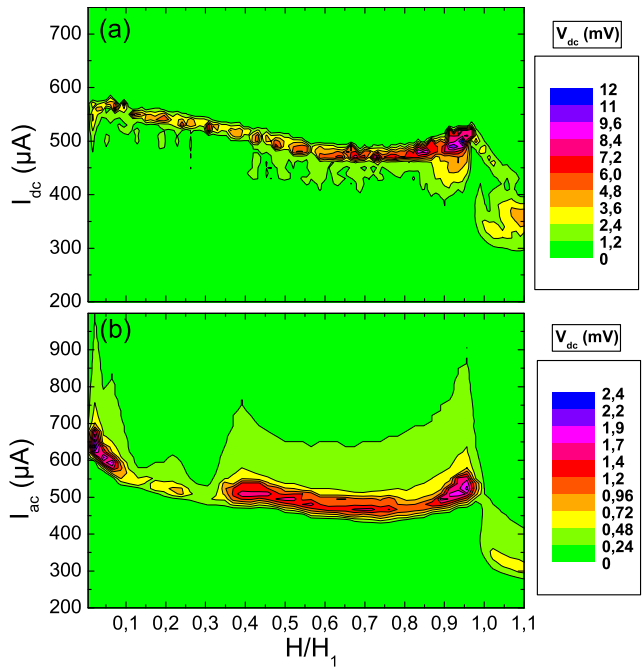}
\caption{(a) Contour plot of the magnetic field and dc current
dependence of the voltage difference for dc currents applied in
the $x$ and $-x$ direction $V(I_{-x},H)-V(I_x,H)$ at $T/T_{c} =
0.973$ (b) Contourplot of the net dc voltage $V_{dc}(I_{ac},
H/H_{1})$ as a function of the magnetic field and amplitude of the
ac sinusoidal current at a frequency 1 kHz and $T/T_{c} = 0.973$.
\label{DC}}
\end{figure}

Fig.~\ref{DC}(a) shows a contour plot of the magnetic field and
current dependence of the difference in induced voltage for the dc
currents flowing in the positive and negative $x$ direction,
$V(I_x,H)-V(I_{-x},H)$, at 97.3 \% of the critical temperature.
The voltage drop in the $x$ direction is a measure of the vortex
velocity in the $y$ direction. Since the $I_{-x}$ and $I_x$ dc
currents induce a Lorentz force in the negative and positive $y$
directions, the vortices will probe the asymmetry of the pinning
sites. We see that for all fields ($0-1.1{H_1}$) there is a dc
voltage difference confirming the presence of an asymmetry in the
pinning potential. This difference is the largest for fields just
below the first matching field, $H_1$, which can be explained by
the high symmetry of the vortex configuration which cancels out
the vortex-vortex interactions. In this case the resulting forces
acting on vortices are coming only from the asymmetric pinning
potential. To get the data shown in Fig.~\ref{DC}(b), a 1 kHz
sinusoidal current was applied to the sample and the output signal
was measured with an oscilloscope used as an analog-digital
converter. Integrating this signal gives the average dc response,
which is plotted as a function of the normalized magnetic field,
$H/H_1$, and amplitude of the applied current, $I_{ac}$. Again, we
can clearly see a ratchet effect at all fields from 0 to
1.1$H_{1}$. This effect is maximum just below $H_{1}$ and
$\frac{1}{2}H_1$, where vortex-vortex interactions are also mostly
cancelled. The measurements carried out at other temperatures
($T/T_{c} = 0.967$ up to $T/T_{c} = 0.98$) show similar features.
Nevertheless, the rectification effect is weaker at higher
temperatures, which can be explained by an increase in $\xi(T)$
and $\lambda(T)$, making the size of the vortices big compared to
the antidot size. In this case the vortices cannot sense the
asymmetry anymore. We have also repeated the measurements for
several frequencies in the range 10 Hz -- 10 kHz. No significant
change in the results was observed, which means that our
experiments are carried out deep in the adiabatic
regime~\cite{Reimann_Rev}.

As it is clearly seen from the vertical scans in Fig.~\ref{DC}(b),
the net dc voltage increases sharply with the ac-current amplitude
up to a maximum value and then decays smoothly to zero. This is a
typical behavior of ac-driven objects in a ratchet potential in
the adiabatic regime. At low frequencies, the net dc flow $\langle
v\rangle$ of particles increases monotonically when the excitation
amplitude $A$ is between the weaker, $f_{d1}$, and the stronger,
$f_{d2}$, depinning forces of the asymmetric potential. We shall
refer to this region as the {\it rectification window}. Since in
this amplitude range the force cannot overcome the potential
barrier for negative force direction, motion occurs only during
the positive half loops of the ac excitation, that is, the system
behaves as a half-wave rectifier. For amplitudes $A>f_{d2}$, the
particles are allowed to travel back and forth, thus leading to a
decrease in the rectification signal, which vanishes slowly at
high drives as the ratchet potential is gradually averaged out by
the fast-moving particles. It is worth noting that these general
properties of the ratchet effect also depend on the shape of the
ac excitation~\cite{Reimann_Rev}. For instance, square-wave
excitations lead to sharper rectification effect and a shorter
tail in the $\langle v\rangle$-$A$ characteristics as compared to
sinusoidal excitations. Since the regime of the dc measurements is
equivalent to irradiating the system with a square wave pulse, the
dc data presented in Fig.~\ref{DC}(a) looks much sharper than the
ac data shown in Fig.~\ref{DC}(b).

\begin{figure}[tb]
\centering
\includegraphics*[width=8.5cm]{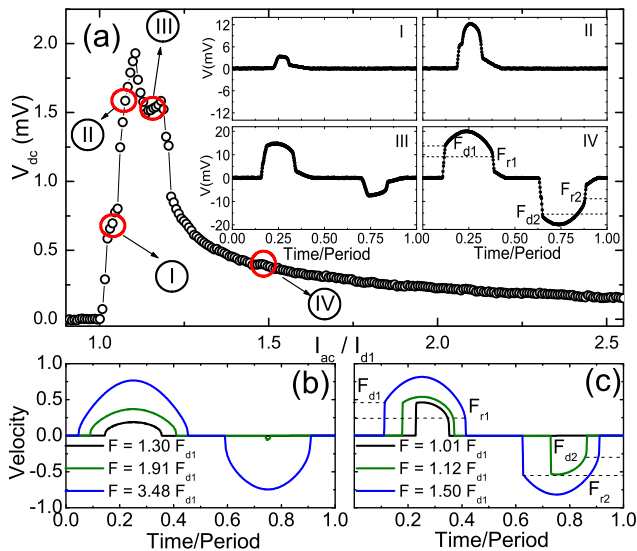}
\caption{Net dc voltage as a function of ac amplitude at $H/H_1 =
0.96$ and $T/T_{c}=0.973$. The current is normalized by the
$I_{d1}=473$ $\mu$A. Inset: Voltage output at the ac amplitudes
(I) 477$\mu$A, (II) 502 $\mu$A, (III) 544 $\mu$A and (IV) 704
$\mu$A; (a) time evolution of the vortex velocity $v(t)$ from an
overdamped molecular dynamics simulation (see text); and (b) v(t)
calculated from the inertia ratchet model. The horizontal dashed
lines indicate the instantaneous value of excitation where
vortices are {\it de}pined [$F(t)=F_d$] and {\it re}pined
[$F(t)=F_r$]. The subscripts 1 and 2 stand for the weaker and the
stronger depinning forces respectively. \label{AC-signal}}
\end{figure}

\begin{figure}[tb]
\centering
\includegraphics*[width=8.5cm]{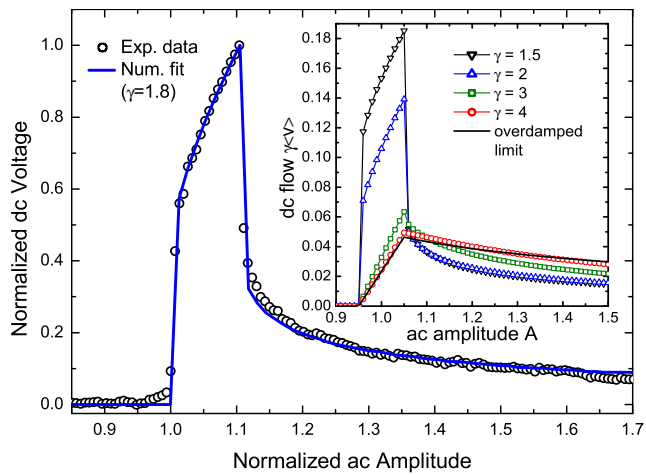}
\caption{\label{Inertia} Normalized dc voltage as a function of ac
amplitude. Circles: experimental data at $H/H_1 = 0.98$ and
$T/T_{c}=0.973$. Line: numerical fit using the inertia ratchet
model [Eq.~(\ref{eq:motion})] with only one fitting parameter,
$\gamma=1.8$. Inset: numerical integration of the inertia ratchet
model for different $\gamma$ (symbols) and the overdamped limit
solution (line).}
\end{figure}

A closer look at the ac-amplitude dependence of the net dc voltage
reveals additional and important features of the ratchet dynamics
in our system. Figs.~\ref{AC-signal} and~\ref{Inertia} show
detailed dc voltage versus ac amplitude ($V_{dc}$-$I_{ac}$)
characteristics. For fields below the first matching field some
$V_{dc}$-$I_{ac}$ curves, as the one at $H=0.96H_{c1}$
(Fig.~\ref{AC-signal}), have two peaks. Detailed measurements of
the time evolution of the output voltage $V(t)$, indicated by
roman numbers, suggest a superposition of two ratchet effects.
Rectification is first triggered just above $I_{ac}=473$ $\mu$A
and the output signal is present only at the positive half-loop of
the ac current, as shown in panel I. At $I_{ac}\simeq 495$ $\mu$A,
a stronger ratchet is triggered. It can be identified in panel II
as a second jump in the signal in the positive half-loop. Right
above $I_{ac}=520$ $\mu$A a small signal in the negative half loop
appears (see panel III), thus demonstrating that the weaker
ratchet is already outside its rectification window. In panel IV,
the signal is strong in both directions, indicating that the
driven vortex lattice moves back and forth as a whole. A possible
reason for the presence of two ratchets can be a complicated
plastic dynamics in which vortex channels flow along specific rows
of pinning centers for currents considerably lower than the
depinning current of the whole vortex lattice. Plastic dynamics
was thoroughly studied by numerical simulations in periodic arrays
of symmetrical pinning centers~\cite{Reichhardt98}. The local
decrease in pinning efficiency in a given pinning row is caused by
discommensurations (vacant pinning sites for fields $H<H_1$)
distributed along the row. The extension of these ideas to arrays
of asymmetric pinning sites seems to be rather straightforward. At
fields very close or at $H_1$ this effect is minimized. At
$H=0.98H_1$ for instance, the $V_{dc}$-$I_{ac}$ characteristic,
shown in Fig.~\ref{Inertia}, presents a single well-defined
rectification peak.

Another remarkable feature disclosed in the $V(t)$ data is that
the repinning force $f_r$ necessary to stop vortex motion in
either direction is always smaller than the corresponding
depinning force $f_d$ [see Fig.~\ref{AC-signal}, insets
(I)--(IV)]. In other words, the V-I characteristics is hysteretic.
This is a very robust phenomenon which has been observed in the
whole field, frequency and temperature ranges used in our
experiment and, to the best of our knowledge, has not been
reported before. To understand the experimental results, we have
performed molecular dynamics (MD) simulations of the overdamped
equations of motion for vortices in a square array of asymmetric
pinning potentials at fields close to the $H_1$ and subjected to a
sinusoidal Lorentz force of very low frequency (the details of the
model and simulations are reported in Ref.~\cite{Zhu}). In the
whole range of excitation amplitudes, no hysteresis was observed
($f_r=f_d$), the $v(t)$ curves at each half period are
symmetrical, in contrast with the clearly asymmetrical shape of
the $V(t)$ experimental data. In addition, the $\langle
v\rangle$-$A$ characteristics look qualitatively different from
the experimental results. A possible reason for the misfit between
the MD calculations and the experimental results is that the
depinning process of an overdamped particle is reversible because
it carries no significative inertia and cannot be deformed.
However, vortices are soft objects that can be deformed under the
action of competing forces and they also carry a small mass. In
the dirty limit, the vortex mass can be several times smaller than
the electron mass. In this case, inertia is only relevant at
frequencies typically above $10^{12}$ Hz~\cite{mass}. Since our
sample is in the dirty limit, we believe the contribution of
vortex mass is negligible in our experiment. Therefore, the
apparent inertia is more likely due to vortex deformation. Indeed,
recent numerical simulations of the Ginzburg-Landau
equations~\cite{Priour03} suggested that the current-induced
depinning of vortices in a periodic array of antidots is dominated
by strong elongation of the vortex cores towards a neighboring
antidot. Interestingly, the authors observed hysteresis in the
pinning-depinning process resulting from this behavior.

To illustrate how this hysteresis influences the overall ratchet
dynamics, we model vortices still as rigid particles but carrying
an apparent mass $M$. For simplicity, we consider that the vortex
lattice is perfectly commensurate with the pinning array and
therefore we shall analyze the experimental results measured very
close to the first matching field ($H=0.978H_1$) shown in
Fig.~\ref{Inertia}. In this case, due to the cancellation of the
vortex-vortex interactions, the vortex dynamics can be reduced to
that of one particle moving in a one-dimensional potential. The
corresponding equation of motion is
\begin{equation} \label{eq:motion}
    M\dot{v} = -\eta v - U_p'(x) + A\sin(\omega t),
\end{equation}
where $\eta$ is the viscous drag coefficient, $v$ is the
instantaneous vortex velocity and $U_p(x)$ is the one-dimensional
pinning potential, which we choose in a simple form as
$U_p(x)=-U[\sin(2\pi x/a_p)+0.025\sin(4\pi x/a_p)]$. This gives an
asymmetry ratio $f_{d1}/f_{d2}=0.925$ similar to that observed in
our data shown in Fig.~\ref{Inertia} ($I_{d1}=480$ $\mu$A and
$I_{d2}=519$ $\mu$A). In order to be closer to the experimental
conditions, we will restrict ourselves to the adiabatic regime,
that is, we shall consider only frequencies much lower than the
libration frequency of the pinning potential $\omega_p\simeq
2\pi\sqrt{U/Ma_p^2}$ and the viscous drag to mass ratio
$\gamma=\eta/M$. In this way, by normalizing the ac amplitude by
the first depinning force and the dc response by maximum
response, the only fitting parameter for comparing the
calculations with the experimental data is $\gamma$.

The overdamped limit corresponds to $M\rightarrow 0$, or,
equivalently, to $2\pi\gamma\gg\omega_p$. In this limit a
semi-analytical approach may be used. The average flow of vortices
in one cycle of the ac excitation is given by $\langle v\rangle =
\frac{1}{P}\int_0^P{\rm d}t\,v(t)$ where $P=\omega/2\pi$ is the
forcing period and
\begin{equation} \label{eq:v_t}
    v(t) = \frac{a_p}{\eta\int_0^{a_p}\frac{{\rm d}x}{-U_p'(x) +
    A\sin(\omega t)}}.
\end{equation}
This model reproduces exactly the MD simulations for $H=H_1$. The
$\gamma\langle v\rangle$-$A$ characteristic calculated for the
overdamped limit is given by the full line in the inset of
Fig.~\ref{Inertia}. The curve presents a long and slowly decaying
tail qualitatively different from the curves observed in our
experiments.

For finite $\gamma$ values, we integrate Eq.~\ref{eq:motion}
numerically by a finite difference method. The results for several
$\gamma$ values (in units of $\omega_p/2\pi$) are shown in the
inset of Fig.~\ref{Inertia}. For $\gamma = 4$, the $\gamma\langle
v\rangle$-$A$ characteristics is already very close to the
overdamped limit. For smaller $\gamma$ values, the effect of
inertia becomes increasingly more pronounced. The curves look
sharper and more restricted to the rectification window. As shown
in the main frame of this figure, the $\gamma=1.8$ curve provides
an excellent fitting to the experimental data. The time evolution
of the velocity $v(t)$ for $\gamma=1.8$ is shown in the inset (b)
of Fig.~\ref{AC-signal} for a few ac drive values. Note that the
curves correctly reproduce the asymmetry observed in the
experimental $V(t)$ data [insets (I)--(IV)], with $F_r\neq F_d$.
The $v(t)$ curves calculated for $\gamma\geq 4$ resemble the
overdamped limit [Eq.~\ref{eq:v_t}], which gives exactly $F_d=F_r$
and no hysteresis.

In conclusion, we have demonstrated voltage rectification in a
superconducting film with a composite periodic array of asymmetric
pinning sites. The rectification was observed in dc and ac
measurements in a broad field range and for different
temperatures, showing that our sample provides a two-dimensional
asymmetric potential which is able to control vortex motion. For
fields below the first matching field, our data revealed a
competition between two ratchet dynamics, which can be understood
as the competition between plastic channelling dynamics and
coherent vortex motion. We also have observed a strong
irreversibility in the pinning-depinning process, which we believe
is due to current-induced elongation of the vortex core.
Calculations based on an underdamped ratchet model provided a very
good fit to the experimental data and showed that the hysteresis
found in our experiment gives rise to an apparent inertia with
important consequences to the ratchet dynamics.

This work was supported by the K.U.Leuven Research Fund
GOA/2004/02 and FWO programs. C.C.S.S. acknowledges the support of
the Brazilian Agency CNPq. M.M. acknowledges support from the
Institute for the Promotion of Innovation through Science and
Technology in Flanders (IWT-Vlaanderen).


\begin{thebibliography}{1}

\bibitem{SecondLaw} M. von Smoluchowski, Phys. Z. {\bf XIII}, 1069
    (1912); R. P. Feynman {\it et al.}, {\it The Feynman Lectures
    on Physics} (Addison-Wesley, Reading, MA, 1963).

\bibitem{Reimann_Rev} For a review see P. Reimann, Phys. Rep. {\bf 361}, 57 (2002).

\bibitem{Linke98} H. Linke {\it et al.}, Appl. Phys. A {\bf 75}, 237 (2002).

\bibitem{Matthias03} S. Matthias and F. M{\"{u}}ller, Nature {\bf 424}, 53 (1999).

\bibitem{Marquet02} C. Marquet {\it et al.}, Phys. Rev. Lett. {\bf 88}, 168301 (2002).

\bibitem{Lee99} C.-S. Lee {\it et al.}, Nature {\bf 400}, 337 (1999).

\bibitem{Wambaugh99}{J.F.~Wambaugh, {\it et al.},
Phys.~Rev.~Lett.~{\bf 83}, 5106 (1999).}

\bibitem{Olson01} C. J. Olson {\it et al.}, Phys. Rev. Lett. {\bf 87}, 177002 (2001).

\bibitem{Zhu} B. Y. Zhu {\it et al.}, Phys. Rev. B {\bf 68}, 014514
(2003); Phys. Rev. Lett. {\bf 92}, 180602 (2004).

\bibitem{Villegas03} J. E. Villegas {\it et al.}, Science {\bf 302}, 1188 (2003).

\bibitem{Wordenweber04} R. W{\"{o}}rdenweber {\it et al.},
Phys. Rev. B {\bf 69}, 184504 (2004).

\bibitem{Reichhardt98} C. Reichhardt {\it et al.}, Phys. Rev. B {\bf 58}, 6534 (1998).

\bibitem{mass} H. Suhl, Phys. Rev. Lett. {\bf 14}, 226 (1965).

\bibitem{Priour03} D. J. Priour and H. A. Fertig, Phys. Rev. B {\bf 67}, 054504 (2003).


\end{thebibliography}
\end{document}